\documentclass[journal=jacsat,manuscript=article]{achemso}

\usepackage[version=3]{mhchem} 
\usepackage{amsmath}
\usepackage{bm}
\usepackage{siunitx}

\author{Evgeniia Slivina}
\affiliation{Institute of Theoretical Solid State Physics, Karlsruhe Institute of Technology, Wolfgang-Gaede-Str. 1, 76131, Karlsruhe, Germany}
\email{evgeniia.slivina@kit.edu}
\author{Derk B{\"a}tzner}
\affiliation{Meyer Burger Research AG Rouges-Terres 61, 2068
Hauterive, Switzerland}
\author{Raphael Schmager}
\affiliation{Institute of Microstructure Technology, Karlsruhe Institute of Technology, Hermann-von-Helmholtz-Platz 1, 76344 Eggenstein-Leopoldshafen, Germany}
\author{Malte Langenhorst}
\affiliation{Institute of Microstructure Technology, Karlsruhe Institute of Technology, Hermann-von-Helmholtz-Platz 1, 76344 Eggenstein-Leopoldshafen, Germany}
\author{Jonathan Lehr}
\affiliation{Light Technology Institute, Karlsruhe Institute of Technology, Engesserstrasse 13, 76131 Karlsruhe, Germany}
\author{Ulrich W. Paetzold}
\affiliation{Institute of Microstructure Technology, Karlsruhe Institute of Technology, Hermann-von-Helmholtz-Platz 1, 76344 Eggenstein-Leopoldshafen, Germany}
\alsoaffiliation{Light Technology Institute, Karlsruhe Institute of Technology, Engesserstrasse 13, 76131 Karlsruhe, Germany}
\author{Uli Lemmer}
\affiliation{Light Technology Institute, Karlsruhe Institute of Technology, Engesserstrasse 13, 76131 Karlsruhe, Germany}
\author{Carsten Rockstuhl}
\affiliation{Institute of Theoretical Solid State Physics, Karlsruhe Institute of Technology, Wolfgang-Gaede-Str. 1, 76131, Karlsruhe, Germany}
\alsoaffiliation{Institute of Nanotechnology, Karlsruhe Institute of Technology, 76344 Eggenstein-Leopoldshafen, Germany}
\title[An \textsf{achemso} demo]
  {The annual energy yield of mono- and bifacial silicon heterojunction solar modules with high-index dielectric nanodisk arrays as anti-reflective and light trapping structures}

\abbreviations{IR,NMR,UV}

\begin{document}







\begin{abstract}
While various nanophotonic structures applicable to relatively thin crystalline sili-con-based solar cells were proposed to ensure effective light in-coupling and light trapping in the absorber, it is of great importance to evaluate their performance on the solar module level under realistic irradiation conditions. Here, we analyze the annual energy yield of relatively thin heterojunction (HJT) solar module architectures when optimized anti-reflective and light trapping titanium dioxide (TiO$_2$) nanodisk square arrays are applied on the front and rear cell interfaces. Our numerical study shows that upon reducing crystalline silicon (c-Si) wafer thickness, the relative increase of the annual energy yield can go up to 11.0\,\%$_\text{rel}$ and 43.0\,\%$_\text{rel}$ for mono- and bifacial solar modules, respectively, when compared to the reference modules with flat optimized anti-reflective coatings of HJT solar cells.
\end{abstract}

\section{Introduction}
Reducing optical losses is of paramount importance for further developing photovoltaic (PV) devices. This holds especially true for the market-dominating single-junction c-Si solar cells, for which optical losses constitute one of the main limitations to reach their efficiency limit~\cite{shockley_detailed_nodate,richter_n-type_2017}. While a common approach involves employing various uniform anti-reflective (AR) coatings combined with a chemical texturing of the c-Si wafer, resulting in the formation of micron-sized pyramidal features~\cite{seidel_anisotropic_1990}, in some cases, alternative approaches to suppress the optical losses are needed. For example, in industrial solar cells, the micron-sized textures are realized on relatively thick c-Si wafers, with a current standard of 160~$\mu$m. However, a transition to the wafer thickness below the standard value and switching to the foil-like thinner c-Si can allow for lower material consumption. This can reduce the manufacturing cost and facilitate the acceleration of the expansion of PV manufacturing~\cite{liu_revisiting_2020} to keep up with estimates for global installed PV capacity~\cite{haegel_terawatt-scale_2019}. However, for such thin devices, the use of micron-sized textures becomes very challenging due to c-Si wafer handling issues, and novel approaches have to be identified to reduce the optical losses.

In response to that need, various nanophotonic concepts~\cite{garnett_photonics_2021} applicable to c-Si-based solar cell stacks were proposed to enhance light harvesting through improved light in-coupling on the front surface. Examples of possible solutions include plasmonic structures~\cite{green_harnessing_2012,Catchpole:08,doi:10.1021/nl8022548}, periodically arranged silicon~\cite{hou_efficient_2020,he_enhanced_2016} and dielectric~\cite{spinelli_effect_2015, saive2020, spinelli_broadband_2012} nanoscatterers, and biomimetic structures~\cite{hou_biomimetic_2017}. Moreover, double-sided AR and light trapping (LT) nanostructure gratings introduced at the front and rear sides of the solar cells were also suggested. For example, such a concept was investigated for thin-film c-Si~\cite{wang_absorption_2012} and thin-film hydrogenated nanocrystalline silicon (nc-Si:H)~\cite{isabella_advanced_2018} solar cells. 

Nevertheless, while a plethora of nanophotonic structures was proposed and investigated in recent years, it is of paramount importance to analyze their performance in a full solar module architecture under realistic irradiation conditions. For example, such analysis was performed for different solar module architectures where sufficient optical properties were achieved using strategies involving textured interfaces and/or flat ARCs~\cite{singh_comparing_2021,C7EE01232B,tucher_energy_2019, lehr_energy_2020,gota2020energy,jovst2018textured}. Since the solar cell is not always illuminated with light at normal incidence in the realistic scenario, one has to go beyond the analysis of the ability of the structure to enhance the short-circuit current density under the standard test conditions and ensure that the proposed AR and/or LT nanostructure designs are robust concerning irradiation impinging on the solar module at increasing angles of incidence. Additionally, the absorption of photons by the solar cell absorber depends on the sun's position. It is influenced by the cloud coverage effect on irradiation received by the module and the module orientation. For solar cells with double-sided photonic nanostructures, it is also appealing to consider and assess the possible power output of the solar modules with nanostructured solar cells in the case of a bifacial module architecture. This module configuration allows for harvesting the photons that can be absorbed when the sunlight either hits the module from its back or is reflected from the ground and can be absorbed in the solar cell. 

Here, we study the energy yield of relatively thin wafer-based c-Si solar modules, for which the solar cell stack is coated with double-sided nanostructure gratings. Suitably designed square arrays of dielectric high-index TiO$_2$ nanodisks are used as AR and LT photonic nanostructures. Their geometrical parameters are subject to optimization.  Results are compared to modules containing more traditional planar thin-film anti-reflective layers. We consider mono- and bifacial modules and introduce the nanodisk array on the front surface of the c-Si-based heterojunction (HJT) solar cell for the former module configuration and both the front and rear side for the latter. The nanodisk arrays are initially optimized at normal incident light by full-wave optical simulations concerning the short-circuit current associated with the reflected portion of the light (details in Sec. \textbf{Calculation of reflectance, transmittance, and absorptance}~\ref{sec:calc}). The optimal design of these nanodisk arrays depends on whether they are employed on the front side, where they serve the purpose of suppressing reflectance, or on the rear side, where they facilitate the light trapping. Therefore, when optimizing the nanodisk arrays, different designs are found depending on the mechanisms through which they contribute to enhancing the absorptance in the c-Si wafer. To estimate the annual energy yield (EY) of the solar module architectures with nanodisk arrays, the optical response from the optimized nanodisk arrays when placed on the front and rear solar cell contact layers is simulated using full-wave optical tools depending on the angle of incidence. This is the primary information fueled afterward into the EY modeling framework. The annual EY is assessed for monofacial and bifacial module architectures with and without TiO$_2$ nanodisk arrays at locations with different climate conditions. The influence of albedo radiation is also considered, which is especially relevant to consider in bifacial module configuration~\cite{russell_influence_2017}. Our key contribution is to show that for the wafer-based c-Si cells with thicknesses for which standard chemical texturing becomes impractical, and alternative AR and LT structures are of interest, the nanodisk arrays that we suggest outperform to a considerable extent the traditional design that relies only on the planar anti-reflective coatings.   
In passing, we note that the design of the nanodisk arrays for front and rear solar cell contacts proposed here is exemplary, and it is neither restricted to a particular solar cell stack nor the materials used. 

\section{Module architectures and numerical methods}\label{methods}

\subsection{Investigated module architectures and material properties}

\begin{figure}[h]
\centering
  \includegraphics[width=\textwidth]{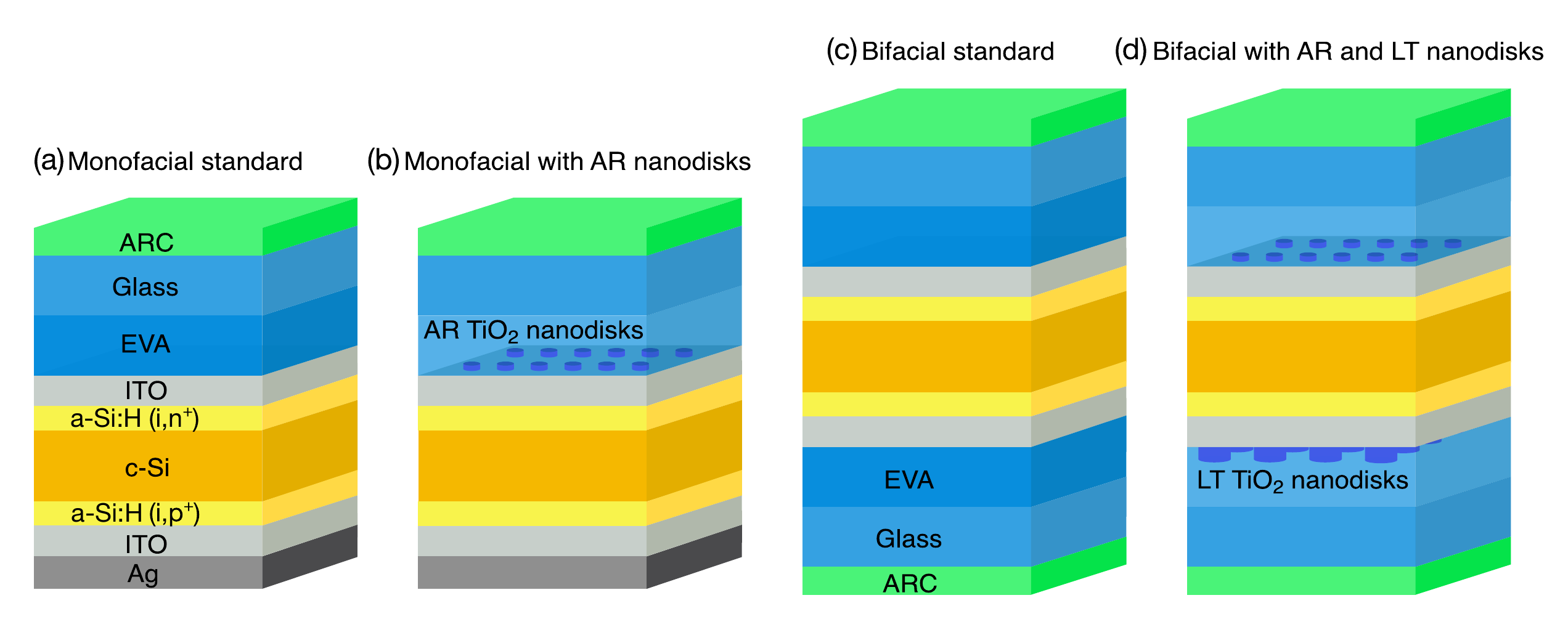}
    \caption{Schematic representation of the four solar module architectures discussed in this paper: (a) Monofacial standard reference module with optimized transparent conductive ITO layer on the front, serving as both anti-reflective coating and front contact, and silver back contact. Front ITO layer is preceded with window glass and encapsulation (EVA) layers covered with an anti-reflective coating. For all considered solar module configurations, the window layers are identical. (b) Monofacial module with optimized anti-reflective nanodisk array on top of the front ITO contact, and silver back contact. (c) Bifacial standard reference module with front and rear ITO contacts with symmetric window layers on both sides of the solar module. (d) Bifacial module with optimized anti-reflective and light trapping nanodisk arrays on top of the front and rear ITO contacts with symmetric window layers on both sides of the solar module, respectively.}
   \label{fig:schematic}
\end{figure}

The four different architectures studied in this contribution are schematically depicted in Fig.~\ref{fig:schematic}. The annual EY calculations were performed for two reference monofacial and bifacial module configurations and two configurations with AR and LT nanodisk arrays introduced on top of the front and rear ITO contact layers, respectively. The c-Si absorber thickness of all architectures was varied between 5 and 80~$\mu$m, which is a typical thickness range between thin PV and conventional wafer-based c-Si PV.

In the case of flat reference c-Si HJT solar cell stacks, the front hydrogenated amorphous silicon (a-Si:H) (passivation intrinsic and n$^{+}$ doped, the thickness can be found in~\cite{slivina_insights_2019}) and conducting ITO (75~nm) layers were considered. For the rear side of the HJT solar cell stack, the a-Si:H (passivation intrinsic and p$^{+}$ doped) layer was slightly thicker than the a-Si:H layer on the front, while the ITO layer was thinner than its counterpart on the front side. 

The configurations containing optimized AR nanodisk array had a reduced front ITO thickness of 10~nm with TiO$_2$ nanodisks arranged in a square lattice of 320~nm pitch, with individual nanodisk having a radius of 125~nm and a height of 90~nm. The bifacial module configuration with added optimized LT TiO$_2$ nanodisk square array (individual nanodisk with a radius of 215~nm and height of 395~nm, 565~nm pitch) had the same rear ITO thickness as the reference architectures. These values for the geometrical parameters are the results of an optimization of the AR and LT nanodisk arrays discussed in Sec. \textbf{Calculation of reflectance, transmittance, and absorptance}~\ref{sec:calc}. 

In the case of the monofacial modules, HJT c-Si solar cell had 300~nm thick silver back contact while the front side contacting metallic grid and its effect on the optical performance of the module is neglected. The window module layers comprise encapsulating EVA (400~$\mu$m) and glass (4~mm) with thin-film anti-reflective MgF$_2$ coating (130~nm). In the case of the bifacial architecture, the rear side of the solar cell typically has the same contacting scheme as the front side (no conformal metallic layer). The modules in the bifacial configuration were considered to have the same window layers as the ones introduced on the front side. In such a configuration, the module can absorb the light that is incident on its front and from its rear side. Additionally, one can harvest albedo radiation since the transmitted portion of the light is reflected from the ground, and thus, can be reabsorbed in the silicon, significantly boosting the annual EY. When a bifacial solar module is tilted, albedo radiation can also impact the annual EY due to the light reflected from the ground and incident of the front of the module. We note that when a monofacial solar module is tilted, albedo radiation that is incident on the front of the module can also be considered.

\begin{figure}[h]
\centering
  \includegraphics[width=7cm]{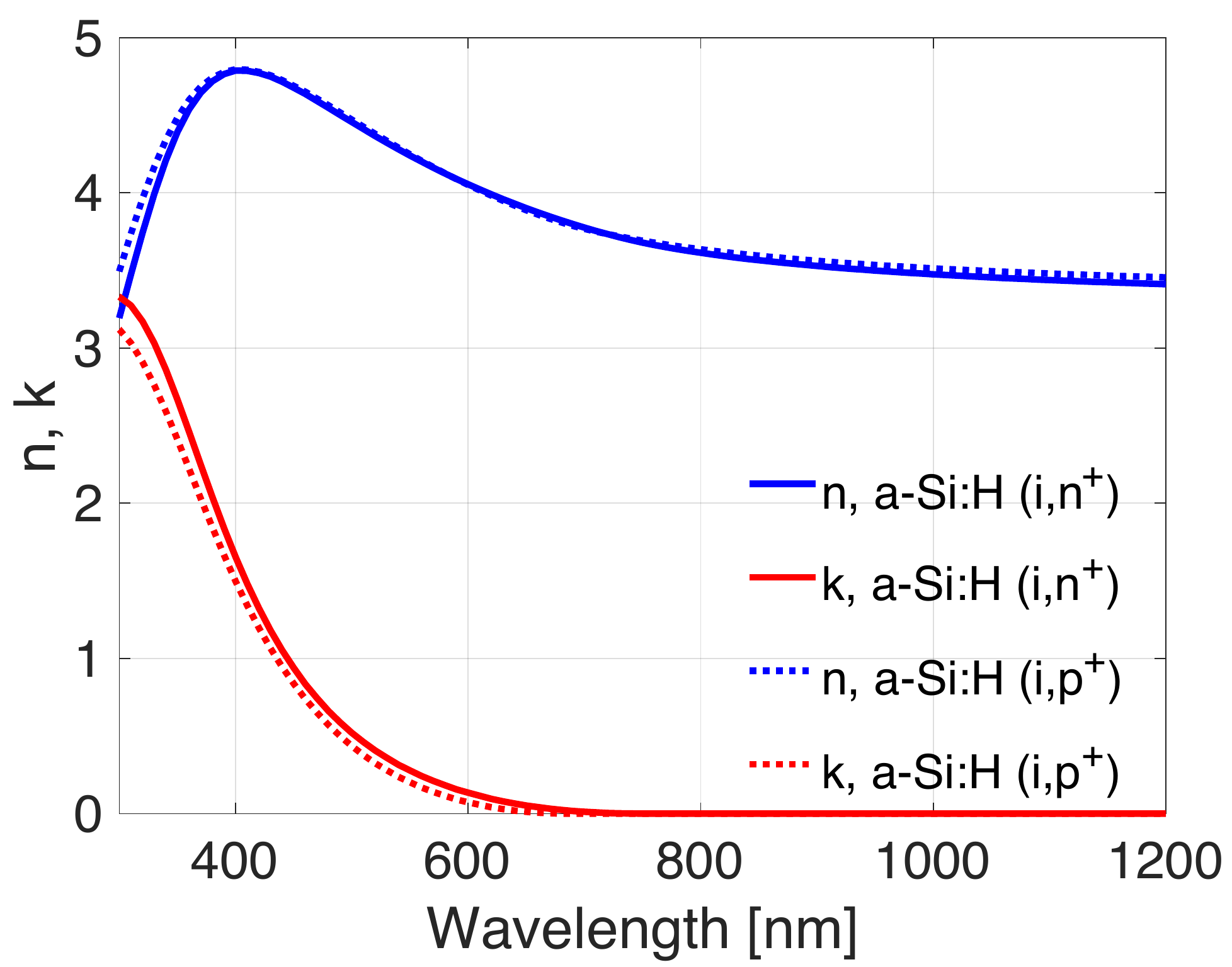}
    \caption{Refractive index $n$ and extinction coefficient $k$ of a-Si:H composite layers. The measured data was fitted to a Tauc-Lorentz model. }
    \label{fig:indices}
\end{figure}

Refractive indices of c-Si, TiO$_2$, ITO, and Ag used in the calculations were taken from literature~\cite{schinke_uncertainty_2015,Rutile,holman_infrared_2013,jiang_realistic_2016}.  Refractive index data for the front and rear composite (passivation intrinsic and doped) a-Si:H layers were obtained using ellipsometry, and corresponding $n$ and $k$ values provided by Meyer Burger Research AG are plotted in Fig.~\ref{fig:indices}. For the window layers, a non-absorbing optically thick glass layer was considered to have a nondispersive refractive index of $n=1.5$, and EVA and MgF$_2$ refractive index data was taken from~\cite{mcintosh_optical_2009} and~\cite{siqueiros_determination_1988}, respectively.
 
\subsection{Simulation framework}
The numerical simulations of the EVA-cell interface for both AR and LT TiO$_2$ nanodisk arrays were performed using the finite element method (FEM) with commercial software \textit{JCMsuite}~\cite{pomplun2007adaptive}. The annual EY was calculated using a comprehensive modeling framework enabling the quick simulation of various and sophisticated PV architectures under realistic irradiation conditions discussed in detail in~\cite{schmager_methodology_2019,GithubEY}.
 
\subsection{Electrical parameters}
The electrical parameters corresponding to a typical c-Si HJT solar cell used in the annual EY calculation are summarized in Table~\ref{tab:electric}. The shadowing by electrical connections for all considered architectures is disregarded. 
 
\begin{table}[h]
\centering
  \caption {Electrical parameters of the solar cell} \label{tab:electric}
    \begin{tabular}{c c}
    \hline
    Shunt resistance, $R_{\text{sh}}$ [$\Omega\cdot\text{cm}^2$] & 5000 \\
    Series resistance, $R_{\text{s}}$ [$\Omega\cdot\text{cm}^2$] & 0.7 \\
    Reverse-blocking current, $J_{\text{0}}$ [$\text{A}/\text{cm}^2$] & 2$\cdot10^{-13}$ \\
    Ideality factor, $n$ & 1.1   \\
    Temperature coefficient of $J_{\text{SC}}$, $t_{J_{\text{SC}}}$ [$\%/\text{K}$] & 0.05 \\
    Temperature coefficient of $V_{\text{OC}}$, $t_{V_{\text{OC}}}$ [$\%/\text{K}$] & -0.25   \\
   \hline
   \end{tabular}
\end{table}

A device characterized by these properties would have a $J_{\text{SC}}$ around 38.3~mA/cm$^2$. This yields $V_{\text{OC}}=0.734$~V at temperature $\text{T}=25^\circ$C from the following equation:
\begin{equation}\label{Voc_RT}
    V_{\text{OC}}=nV_\text{th}\text{ln}\left(\frac{J_{\text{SC}}}{J_{\text{0}}}+1\right),
\end{equation}
where the thermal voltage $V_\text{th}=kT/q=0.0257$~V, and the values of ideality factor and reverse-blocking current can be found in Table~\ref{tab:electric}.
 
\subsection{Calculation of reflectance, transmittance, and absorptance}\label{sec:calc}
 
At the EVA-cell interfaces with introduced AR and LT nanodisk arrays, reflectance and transmittance into all scattering directions at each wavelength and incidence polar and azimuth angles is calculated as the ratio of the scattered reflected or transmitted power to the power of the incident field. At the considered interfaces, both c-Si and EVA are assumed to be semi-infinite. For a given azimuth angle $\phi_\text{in}$, reflectance and transimittance values form matrices of the size ($N_{\theta_\text{in}}$, $N_{\theta_\text{r,t}}$, $N_{\lambda}$), where the entries correspond to all polar angles of incidence, scattering angles, and wavelengths, respectively. The polar angle $\theta_\text{in}$ is varied from $\ang{0}$ to $\ang{89}$ with $\ang{5}$ step, and the results are then interpolated at intervals of $\ang{1}$. In case of azimuth angle $\phi_\text{in}$, the symmetry of the nanodisk coating is exploited, and only calculations for angles between $\ang{0}$ and $\ang{45}$ with $\ang{15}$ step are performed. The calculated matrices for different $\phi_\text{in}$ values are subsequently averaged. Total reflectance and transmittance for a certain wavelength and incident polar and azimuth angles are calculated according to: 
\begin{equation}\label{eq:Refl}
    R=\frac{\sum\limits_{\bm{k}_\text{r}}|\Tilde{\bm{E}}(k_{\text{r,x}},k_{\text{r,y}})|^2\cdot \text{cos}(\theta_\text{r})}{|\bm{E}_0|^2\text{cos}(\theta_\text{in})},
\end{equation}
\begin{equation}\label{eq:Trans}
    T=\frac{\sum\limits_{\bm{k}_\text{t}}n_{\text{out}}|\Tilde{\bm{E}}(k_{\text{t,x}},k_{\text{t,y}})|^2\cdot \text{cos}(\theta_\text{t})}{n_{\text{in}}|\bm{E}_0|^2\text{cos}(\theta_\text{in})},
\end{equation}
where $\bm{k_\text{r,t}}$ are the wave vectors of reflected and transmitted fields, $\theta_\text{r,t}=\Re{(\pm k_\text{z}/k_\text{r,t})}$ are the scattering angles, $\bm{E}_0$ is the amplitude of an incident plane wave with a mixed TE-TM polarization, and $n_\text{in}$ and $n_\text{out}$ are the refractive indices of the media where the incident and scattered waves propagate, respectively. We use the angular spectrum representation of the fields, and $\Tilde{\bm{E}}(k_{\text{\{r,t\},x}},k_{\text{\{r,t\},y}})$ in equations~\ref{eq:Refl} and~\ref{eq:Trans} is calculated by means of the Fourier transform of the electric fields in real space obtained from full-wave simulations. Absorptance in each of the thin film layers and nanodisk array was calculated by integrating the divergence of the Poynting vector across the absorber volume, thus yielding absorbed power which is normalized to the power of the incident plane wave. Similarly to reflectance and transmittance, for a given azimuth angle, absorptance values form a matrix of the size ($N_m$, $N_{\theta_\text{in}}$, $N_{\lambda}$), where index $m$ runs over all absorbing layers in the front or rear solar cell stack. The angular dependent simulations were performed for optimized AR and LT nanodisk arrays. The optical performance of the nanodisk arrays was optimized with respect to the short-circuit current associated with reflectance at a normal light incidence, which is calculated using the following equation:
\begin{equation}\label{eq:JscR}
    J_{\text{SC,R}}=\int_{\lambda_1}^{\lambda_2}e \frac{SI_\text{AM1.5}(\lambda)R(\lambda)}{E_\text{ph}}d\lambda,
\end{equation}
where $e$ is the electron charge, $E_\text{ph}=hc/\lambda$ is the energy of a photon, and $SI_\text{AM1.5}(\lambda)$ is the spectral irradiance. For this calculation, air mass 1.5 global (AM1.5G) tilted irradiance raw data was taken from~\cite{gueymard1995smarts2}, and the total reflectance $R(\lambda)$ was interpolated accordingly. The short-circuit current due to reflectance $J_{\text{SC,R}}$ was minimized for front  nanodisk array and maximized for the rear LT nanodisk array. 

Within the EY modeling framework, where the optical response of the entire architecture is computed, the light propagation in multi-layer thin-film stacks is treated coherently, for which the transfer matrix method is employed. When AR and LT nanodisk arrays are considered instead of those thin-film layers, the corresponding output matrices for reflectance, transmittance, and absorptance are integrated into the modeling framework. For thicker layers, such as the c-Si substrate of the cell and window layers of the module, the assumption of coherence breaks down. The Beer-Lambert law can describe the absorption of the light in those layers:
\begin{equation}\label{eq:Beer}
    I(z,\lambda)=I_\text{0}\cdot\text{e}^{-\alpha(\lambda)z},
\end{equation}
where $I_\text{0}$ is the initial intensity, $\alpha$ is the absorption coefficient of the considered medium, and $z$ is the distance traveled in it.

\section{Results and discussion}

\subsection{Energy yield of solar modules}

We analyzed the annual EY of the four solar module architectures introduced in Fig.~\ref{fig:schematic} for three cities in the United States of America located in different climate zones~\cite{peel2007updated}. Two of the chosen cities, Anchorage, AK and Honolulu, HI, have highly contrasting irradiation conditions. The former is a cold and cloudy region (Boreal climate) and the latter a hot and sunny one (Tropical climate). The additionally chosen Kansas City, MO, has a temperate climate that receives an annual solar irradiance between Anchorage and Honolulu. By covering different climate zones, we aimed to highlight the robustness of the nanodisk arrays performance and their ability to improve the annual EY for all types of irradiation conditions, albeit with small differences that most likely originate from the spectral features of the nanodisk arrays. The solar modules were considered to face south, and the tilt angles $\theta_\text{m}$ were optimized for each location. This resulted in $\theta_\text{m}$ values to be $\ang{38}$ for Anchorage, $\ang{30}$ for Kansas City, and $\ang{17}$ for Honolulu, respectively. 

\begin{figure}[h]
\centering
  \includegraphics[width=13cm]{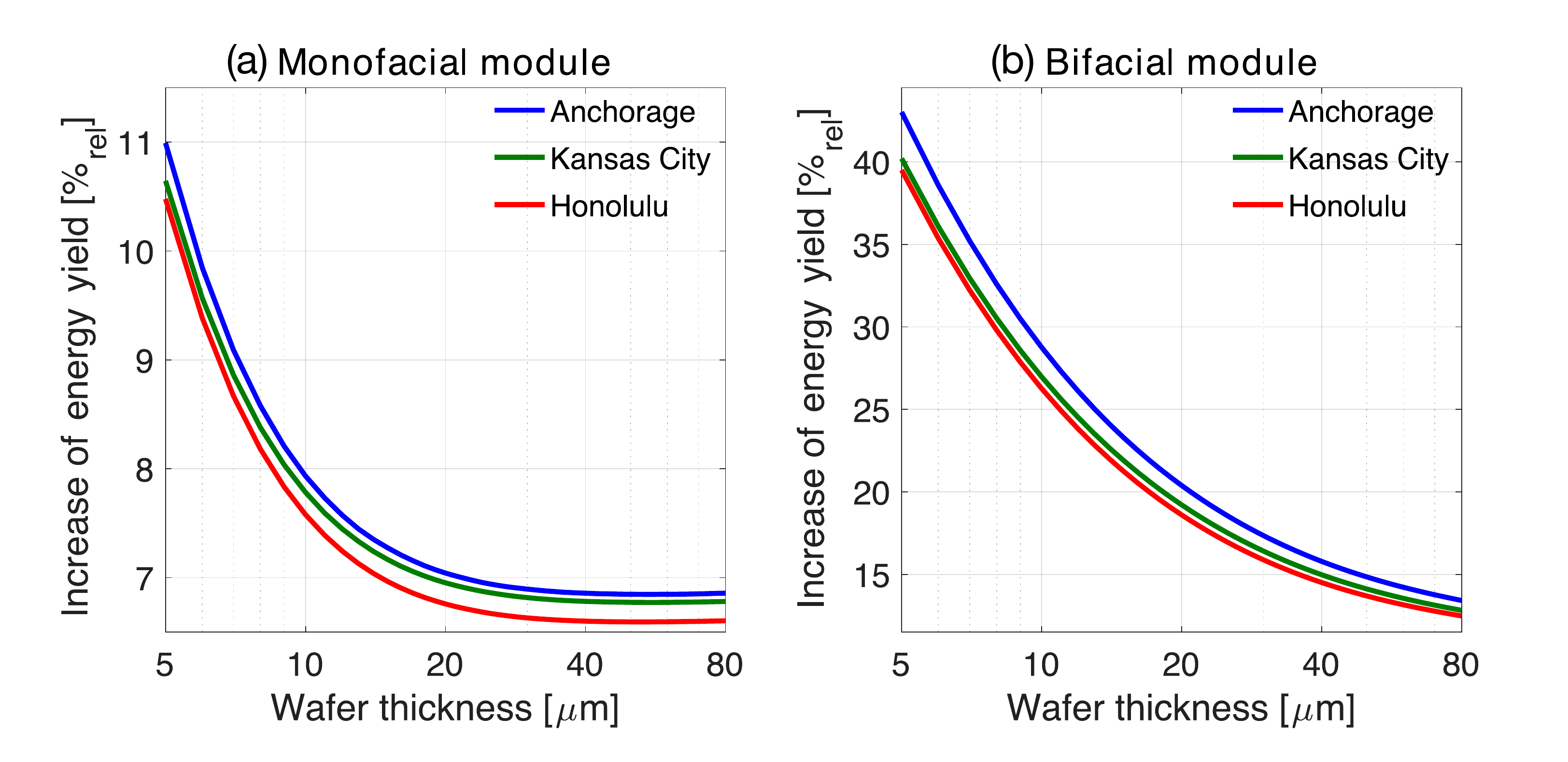}
    \caption{Relative increase of the annual EY for three locations in case of (a) the monofacial solar module (comparing (b) to (a) from  Fig.~\ref{fig:schematic}) and (b) the bifacial solar module (comparing (d)  to (c) from Fig.~\ref{fig:schematic}) with varying thickness of the c-Si absorber.}
   \label{fig:changeEY}
\end{figure}

Figure~\ref{fig:changeEY} demonstrates the relative improvement of the annual EY when the nanodisk arrays are used for light management instead of the optimized planar layers. The increase of the annual EY is shown as a function of the c-Si absorber thickness. We considered the case when the AR array is introduced on the front surface of the c-Si HJT cell in the case of the monofacial module architecture, and both AR and LT nanodisk arrays are applied to the front and rear surfaces of the cell in the case of the bifacial module architecture. For this calculation, no albedo was considered.  The relative increase of the annual EY reached up to 11.0\,\%$_\text{rel}$ and 43.0\,\%$_\text{rel}$ at the minimal wafer thickness of 5~$\mu$m for monofacial and bifacial architectures with nanodisks, respectively. 

As expected, for the monofacial case, the module with an AR nanodisk array (Fig.~\ref{fig:schematic}(b)) outperforms the standard flat architecture with an optimized ITO coating (Fig.~\ref{fig:schematic}(a)), with this effect becoming even more apparent when reducing the c-Si absorber thickness. However, the front side AR array with a relatively small size of the individual disks does not have strong LT properties. Its contribution to light harvesting stems mainly from its broadband AR performance. When instead of a full metallic contact, the window encapsulation and glass layers are introduced, to take advantage of the solar irradiation which can hit the module on its back, one has to take care of an appropriate light trapping structure, which would also act as a decent AR coating. For this purpose, an LT nanodisk array was designed (Fig.~\ref{fig:schematic}(d)). With the individual disk parameters being significantly larger when compared to the nanodisk array used as the AR structure at the front interface, the light which is not absorbed in the silicon and reaches the rear interface of the cell is effectively scattered in multiple directions, thus improving the LT properties of the cell. From Fig.~\ref{fig:changeEY} (b) it can be seen that it translates into an even more significant increase of the annual EY than for monofacial module architecture when comparing to a reference bifacial module architecture (Fig.~\ref{fig:schematic}(c)).

\begin{figure}[h]
\centering
  \includegraphics[width=\textwidth]{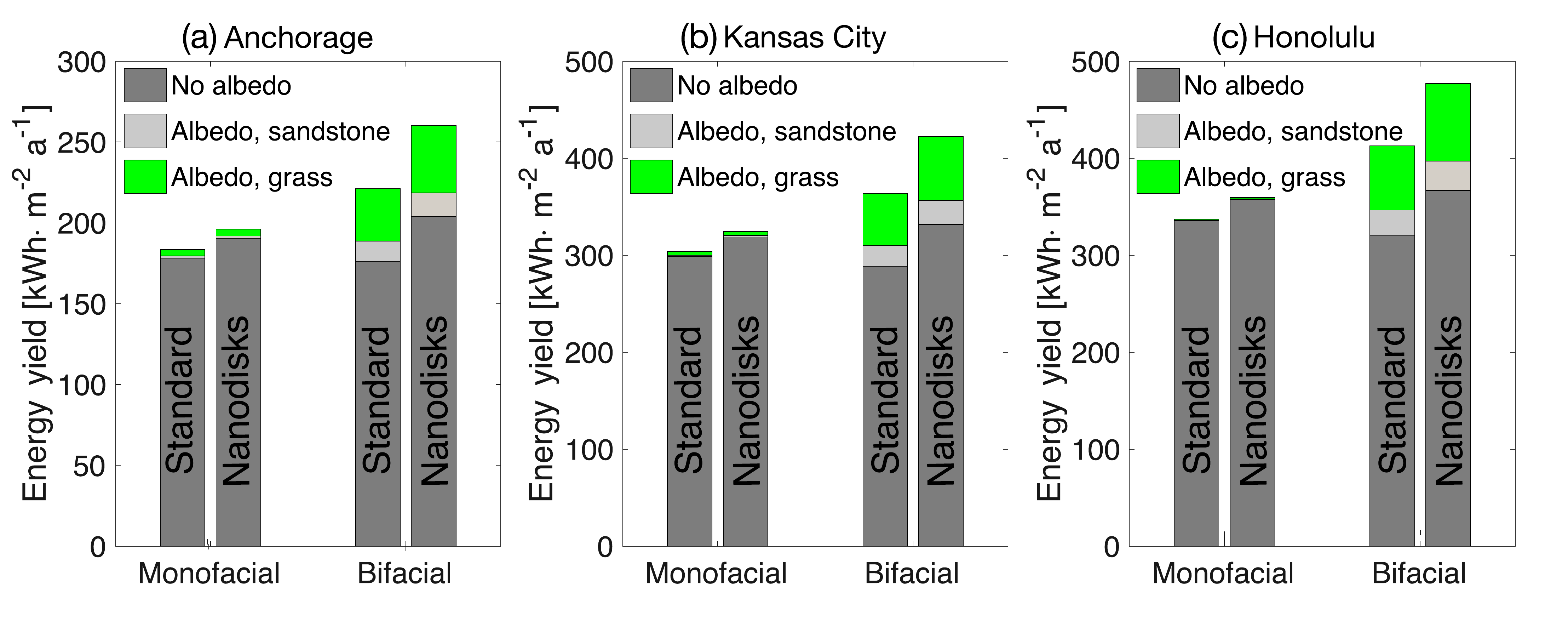}
    \caption{Energy yield of the four module architectures for different locations when albedo irradiation is taken into account. For all solar module architectures, the c-Si absorber thickness was 40~$\mu$m.}
    \label{fig:barEY}
\end{figure}

If additionally, one considers albedo radiation, the advantage of the bifacial module architecture with AR and LT nanodisk arrays becomes even more apparent. The influence of albedo is shown in Fig.~\ref{fig:barEY}. Here, the annual EY of the four module architectures with a selected median c-Si absorber thickness of 40~$\mu$m is shown for the sandstone and grass ground surface compared to EY of the modules without albedo.
Since the interfaces of the standard reference modules are flat, without albedo, the monofacial standard reference module outperforms the bifacial reference. While for the former, the silver layer reflects the light reaching it back into the cell, for the latter, a lot of light is lost due to transmittance when no LT structure is introduced. This difference is bigger with a smaller module tilt angle since less irradiation can hit the module from the back. However, as soon as albedo radiation is considered, the bifacial standard reference outperforms its monofacial counterpart. While for monofacial architecture albedo radiation does not make a significant difference, one can see a robust improvement in the annual EY for the bifacial case. The increase of the annual EY for the monofacial architecture varies depending on the module tilt and is stronger for locations with a greater $\theta_\text{m}$ (Anchorage). The relative improvement reaches up to around 0.8\,\%$_\text{rel}$ with sandstone and 3.0\,\%$_\text{rel}$ with grass as a ground surface for both the standard reference monofacial module and monofacial module with AR nanodisk array with insignificant difference between them. In the case of the bifacial module architecture, a stronger increase of the annual EY is expected for the sunnier locations (Honolulu). It reaches up to 8.2\,\%$_\text{rel}$ with sandstone, and 28.9\,\%$_\text{rel}$ with grass ground surface in case of a standard reference bifacial module. For a bifacial module with AR and LT structures, the relative increase is 8.3 \,\%$_\text{rel}$ and 30.1\,\%$_\text{rel}$ with sandstone and grass ground surface, respectively.     

\subsection{Optical performance of solar modules}\label{sec:opt}

\begin{figure}[h]
\centering
  \includegraphics[width=12cm]{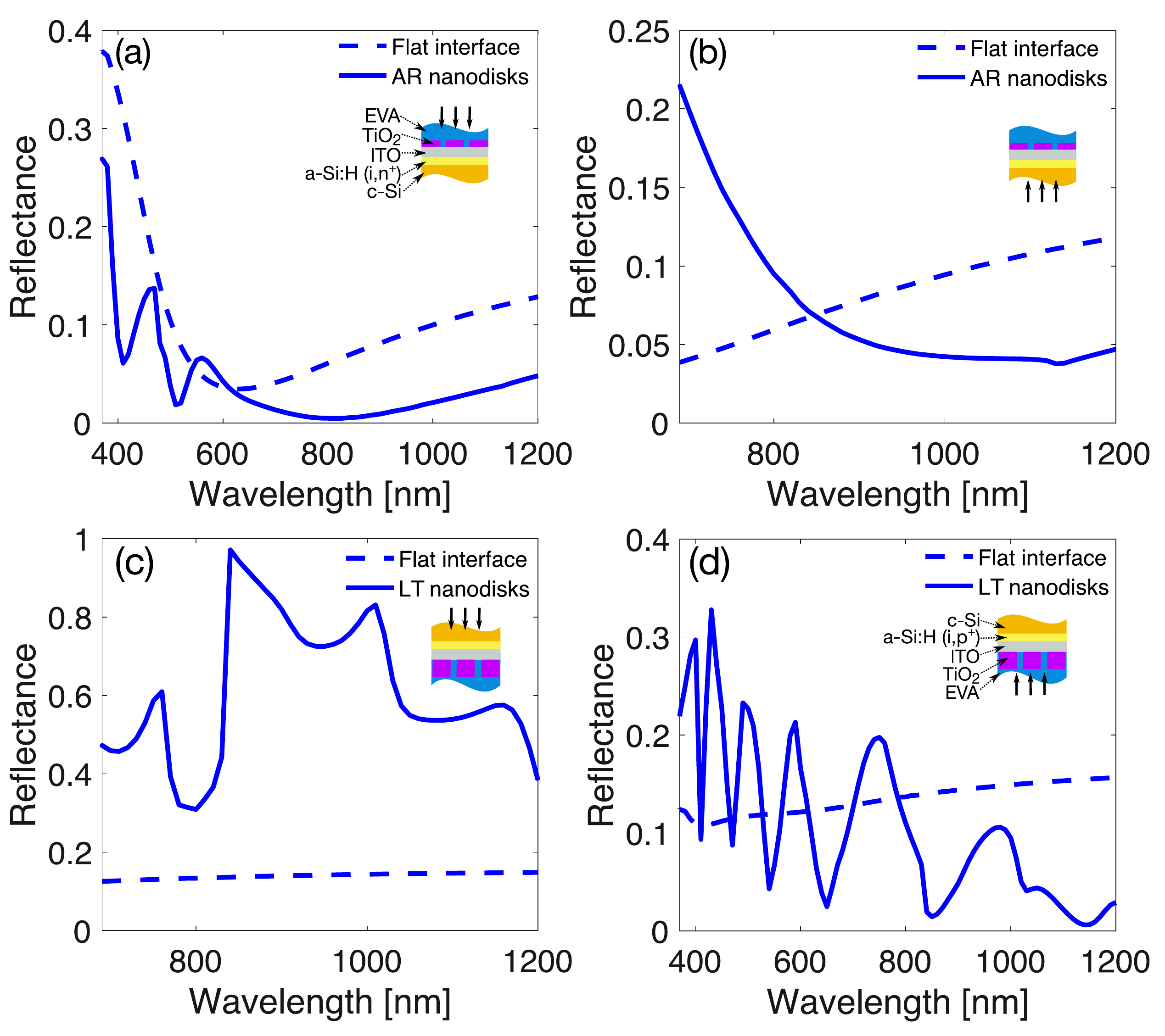}
    \caption{Reflectance at normal incidence for the front and rear EVA-cell interfaces in case of both AR and LT nanodisk arrays. Inset sketches depict the corresponding interface and illumination direction. Interfaces are the same for the pairs of graphs (a)-(b) and (c)-(d), respectively.}
    \label{fig:Reflectance}
\end{figure}

To better understand the mechanism behind the annual EY improvement when the nanodisk arrays are introduced on top of the front and back contacts of a solar cell, one can look at the optical properties of the cell interfaces with the optimized nanodisk arrays. Figure~\ref{fig:Reflectance} shows reflectance at normal incidence of the EVA-cell interfaces for the front and rear HJT solar cell contacts with AR and LT nanodisk arrays. For both nanodisk arrays and illumination directions, reflectances of the corresponding reference flat interfaces are plotted for comparison. The graphs (a) and (b) in Fig.~\ref{fig:Reflectance} correspond to the optical response of the front EVA-cell interface. In this case, the objective was to minimize short-circuit current density corresponding to reflectance $J_{\text{SC,R}}$ introduced in Eq.~\ref{eq:JscR}, which resulted in an AR nanodisk array with reflectance shown in Fig.~\ref{fig:Reflectance}(a). 

This design was based on the previous work on helicity preserving TiO$_2$ nanodisk array for the front interface of a c-Si HJT solar cell, where efficient and broadband backscattering suppression was achieved due to the ability of the system to suppress cross-talk between opposite handednesses of the electromagnetic field upon light-matter interaction~\cite{slivina_insights_2019}. The requirement for a system to be helicity preserving is to possess a high enough degree of rotational symmetry ($n\ge3$) along the illumination direction. For normal light incidence, for which the optimization of AR nanodisk array was performed, the illumination direction is along the symmetry axis of an individual nanodisk, which essentially means that $n\rightarrow\infty$ in this case. The resulting reflectance of the AR nanodisk array is lower than the one of the reference optimized flat AR coating and exceeds it slightly only in the wavelength region around $\lambda=600$~nm, for which standard AR coating of the solar cell is typically optimized. However, as shown in Fig.~\ref{fig:Reflectance}(b), the LT properties of this nanodisk array are not as good as its AR properties. When the light is impinging from the c-Si absorber, only the long-wavelength response is relevant since the short-wavelength photons are absorbed before reaching this interface. This nanodisk coating transmits the light reflected from the rear of the stack particularly strongly at longer wavelengths and is not possessing better LT properties than the standard flat reference. This optical response confirms that the front nanodisk array contributes to the light harvesting of the considered solar module architectures mainly by its AR properties. 

The graphs (c) and (d) in Fig.~\ref{fig:Reflectance} demonstrate the optical response of the optimized LT nanodisk array. Here, the structure parameters strongly differ from those of the helicity preserving AR nanodisk array. The larger and more sparsely spaced LT nanodisks allow for improved harvesting of the long-wavelength photons reaching the rear of the solar cell. The impinging light is effectively scattered into multiple directions since many diffraction orders are allowed. Taking into account the absorption depth of c-Si, for the minimal considered absorber thickness of 5~$\mu$m, the photons that can reach the rear contact of the solar cell have wavelengths $\lambda\geq690$~nm. The optimized reflectance of the LT nanodisk array for the spectral region of interest is shown in Fig.~\ref{fig:Reflectance}(c). Its LT performance exceeds that of the flat reference solar cell stack at all wavelengths. Nevertheless, as can be seen from  Fig.~\ref{fig:Reflectance}(d), though this nanodisk array outperforms its flat reference counterpart in terms of AR properties in the longer wavelength range, overall, its main contribution to the improved absorption in c-Si, and, consequently, the annual EY of the solar module, is due to its superior LT properties. This way, in the bifacial module case, when nanodisk arrays with decoupled AR and LT properties are present on both sides of the solar cell stack, one can achieve a solid overall solar module performance boost. 

\begin{figure}[h]
\centering
  \includegraphics[width=12cm]{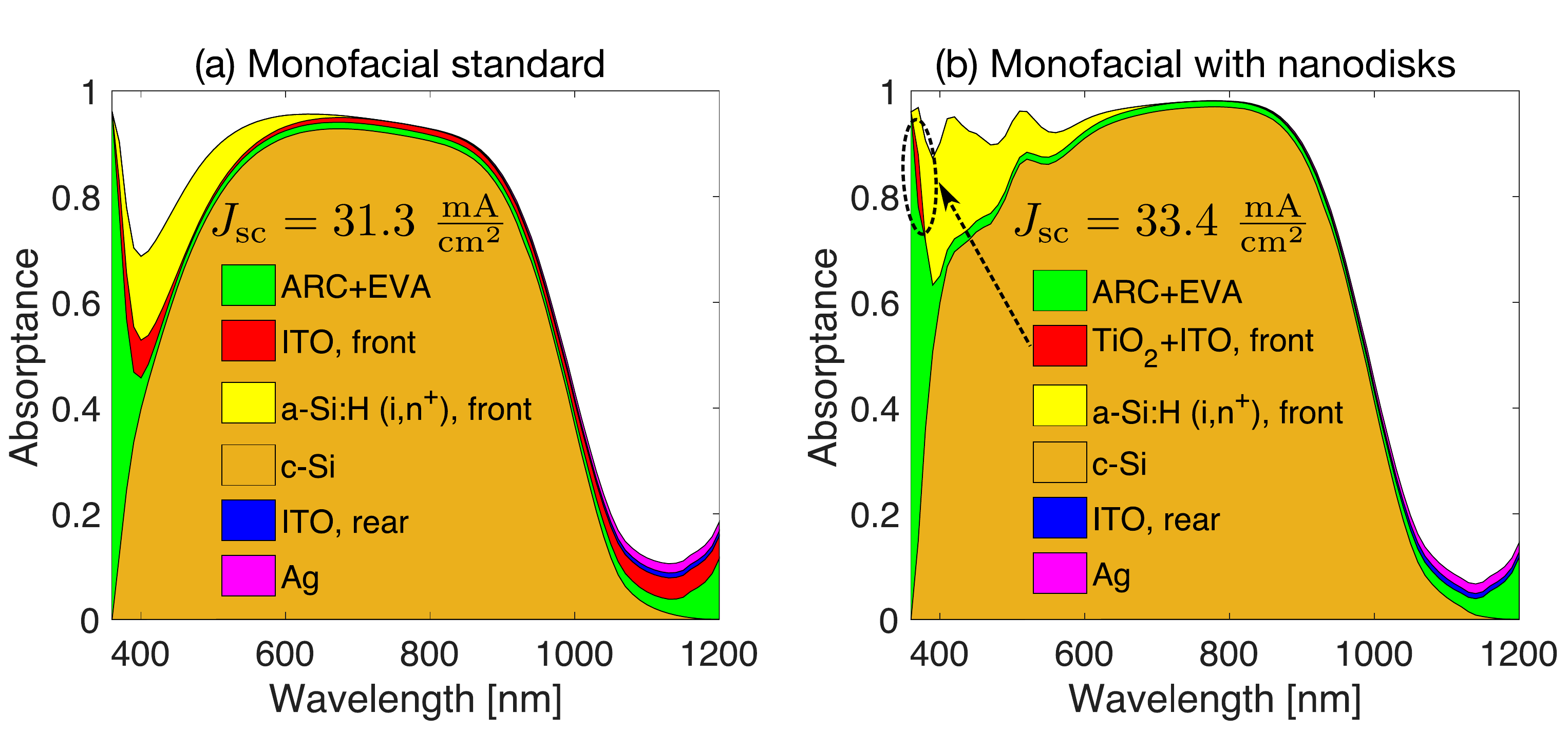}
    \caption{Absorptance in the different layers of (a) the monofacial standrad reference module and (b) the monofacial module with AR nanodisk array on top of the front ITO contact at normal light incidence. For both solar module architectures, the c-Si absorber thickness was 40~$\mu$m.}
    \label{fig:AMono}
\end{figure}

\begin{figure}[h!]
\centering
  \includegraphics[width=12cm]{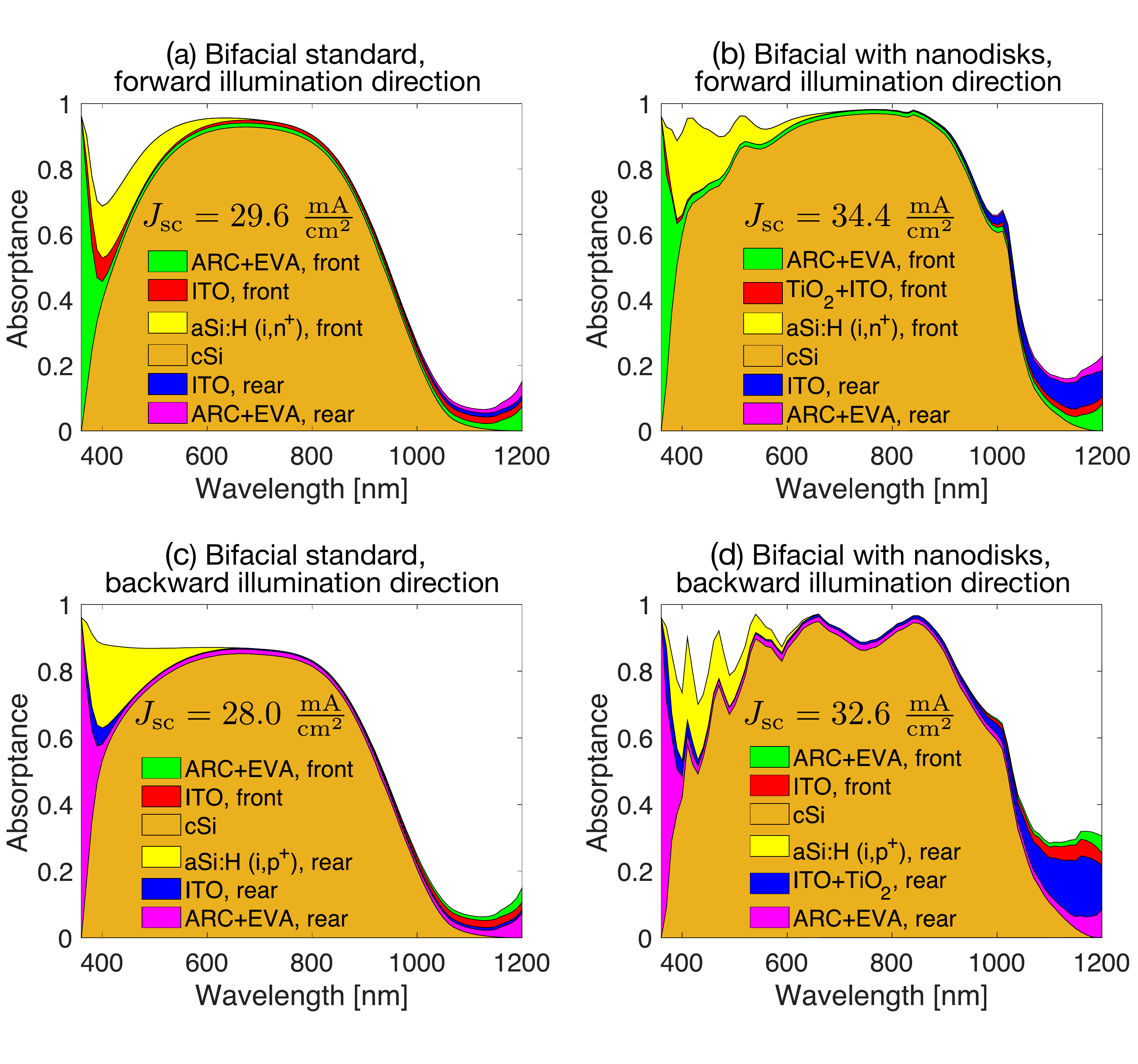}
    \caption{Absorptance in the different layers of the bifacial solar modules  at normal light incidence: (a) in case of forward illumination direction for the bifacial standard reference module, (b) in case of forward illumination direction for the bifacial module with AR and LT nanodisk coatings, (c) in case of backward illumination direction for the bifacial standard reference module, (d) in case of backward illumination direction for the bifacial module with AR and LT nanodisk coatings. For both solar module architectures, the c-Si absorber thickness was 40~$\mu$m.}
    \label{fig:ABi}
\end{figure}

Another critical aspect of the solar module's optical performance is parasitic absorption. Here, all discussed results are for a selected median c-Si absorber thickness of 40~$\mu$m and at normal light incidence. Figures~\ref{fig:AMono} and~\ref{fig:ABi}  show absorptance in all layers of the monofacial and bifacial module stacks, respectively, except for the glass layer, which was assumed to be non-absorbing. Additionally, the absorptance of the rear a-Si:H layer in monofacial module configuration and of rear and front a-Si:H layers in the bifacial module configuration for forward and backward illumination direction, respectively, was negligible, and, thus, it is not shown. Moreover, the LT TiO$_2$ nanodisk array, which is considered for the bifacial module configuration, does not introduce any parasitic absorption since the light absorbed in these nanodisks has short wavelengths and does not reach the rear cell interface while it is absorbed in c-Si. The short-circuit current densities indicated on the top of all graphs in Fig.~\ref{fig:AMono} and Fig.~\ref{fig:ABi} were calculated assuming AM1.5G spectrum using the modified Eq.~\ref{eq:JscR} with absorptance of c-Si instead of reflectance. 

In the case of the monofacial standard reference module (Fig.~\ref{fig:AMono}(a)), the front transparent conductive ITO layer serves as an AR coating but also introduces some parasitic absorption. However, when the AR nanodisk array is introduced (Fig.~\ref{fig:AMono}(b)), ITO thickness for the optimized front solar cell contact is reduced considerably (from 75 to 10~nm), and parasitic absorption in this layer is significantly reduced. The nanodisks themselves have a tiny contribution to the parasitic absorption, considering that EVA absorbs all the light impinging on the solar module at the wavelengths between 300 and 360~nm and TiO$_2$ absorbs only at wavelengths shorter than $\lambda=380$~nm. On the other hand, while the AR nanodisk array increases the absorptance in c-Si thanks to a better in-coupling of the incident light, it also increases parasitic absorption in the front a-Si:H layer. However, it should be noted that in~\cite{holman_current_2012} it was shown that the carriers which are absorbed in an intrinsic a-Si:H layer can still contribute to the short-circuit current, and thus the parasitic absorption loss in this layer represents an upper bound and can have a less of an impact in reality. Thus, introducing the TiO$_2$ AR nanodisk array leads to a broadband enhancement of absorptance in the silicon absorber layer.  

In the case of the bifacial standard reference module and forward illumination direction (Fig.~\ref{fig:ABi}(a)), the parasitic absorption for the wavelengths below $\lambda=700$~nm is similar to the one of the monofacial module reference. However, the parasitic absorption is slightly lower in the long-wavelength range since more light is transmitted through the glass and encapsulation window layers. When AR and LT nanodisk arrays are introduced (Fig.~\ref{fig:ABi}(b)), the front nanodisks improve absorptance in silicon the same way as in the case of the monofacial module. The rear nanodisk array additionally enhances aborptance at longer wavelengths. However, due to the strong scattering enabled by LT nanodisks, parasitic absorption is also increased for longer wavelengths. When the light is impinging on the rear side of the bifacial solar module (Fig.~\ref{fig:ABi}(d)), the absorptance in c-Si is also improved in the case when AR and LT arrays are introduced in comparison to the reference flat module (Fig.~\ref{fig:ABi}(c)), even though LT nanodisks are not optimal in terms of their AR properties and introduce dips in silicon absorptance due to the sharp spectral features which can be seen in Fig.~\ref{fig:Reflectance}(d). 

\section{Conclusions}
We have numerically studied the annual energy yield (EY) under realistic irradiation conditions for monofacial and bifacial crystalline silicon (c-Si) heterojunction (HJT) solar module architectures with AR and light trapping (LT) titanium dioxide (TiO$_2$) nanodisk square arrays introduced on top of the front and rear ITO layers and compared their power outputs with the ones of the corresponding standard reference flat solar modules. We have shown that while reducing the silicon absorber thickness, the relative increase of the annual EY is reaching up to 11.0\,\%$_\text{rel}$ and 43.0\,\%$_\text{rel}$ for monofacial and bifacial modules with nanodisk coatings, respectively. This improvement is comparable for the locations with different climate conditions. Moreover, in the case of bifacial module architecture, taking into account the albedo radiation produces an additional boost of the module performance.  

The designed dielectric nanodisk arrays for the front and rear contacts of c-Si HJT solar cell have both a significant impact on the light absorption in the c-Si wafer. At the same time, their AR and LT properties are decoupled. The front AR nanodisk array has a relatively small individual disk size and lattice constant, and its broadband backscattering suppression is related to the helicity preservation condition. In contrast, the rear LT nanodisk array has larger features and array pitch and allows for efficient scattering into multiple scattering directions. Furthermore, these AR and LT nanodisk square array designs are not restricted to a specific material or a particular photovoltaic solar cell stack and, thus, can be investigated for different solar module configurations. 

\begin{acknowledgement}

The authors acknowledge support by the state of Baden-W{\"u}rttemberg through bwHPC and the German Research Foundation (DFG) through grant no INST 40/575-1 FUGG (JUSTUS 2 cluster). The authors are also grateful to JCMwave for their free provision of the FEM Maxwell solver JCMsuite. 

\end{acknowledgement}




\bibliography{achemso-EY}

\end{document}